# Schwarzite and schwarzynes based load-bear resistant radial cellular griding-based 3D printed structures

Eliezer F. Oliveira[a,b,†], Rushikesh S. Ambekar[c,†], Douglas S. Galvao[a,b,*], Chandra S. Tiwary[c,*]

[a]Applied Physics Department, State University of Campinas (UNICAMP), Campinas, SP, Brazil

[b]Center for Computational Engineering & Sciences (CCES), State University of Campinas - UNICAMP, Campinas, SP, Brazil

[c]Metallurgical and Materials Engineering, Indian Institute of Technology Kharagpur, Kharagpur-721302, India

[†] Equal contribution

*Email: chandra.tiwary@metal.iitkgp.ac.in (CST) and galvao@ifi.unicamp.br (DSG)

**Abstract**

Nature-occurring structures exhibiting unique topological features such as complex and gradient porosity has been the basis to create new materials and/or structures. Most studies have been focused on complex periodic porous structures but gradient porous ones have not been yet fully investigated for stable structural designs. In this work, we have proposed and tested a new approach to create cellular griding structures, in which the mass density varies from the center to the borders, *i.e*, a radial gradient. To create these new structures we exploited the topology of two carbon-based families with different pore sizes, the schwarzites, and schwarzynes. We created fully atomistic models that were translated into macroscale ones that were then 3D printed. The mechanical behavior of the gradient structures was investigated by molecular dynamics simulations and mechanical compression tests of the printed models. Our results show that their mechanical response can be engineered (for instance, in terms of energy absorption, ballistic performance, etc.) and can outperform their corresponding density uniform structures.

# 1. Introduction

Nowadays, there is an increasing demand for lightweight materials/structures that could result in the development of new products for several applications, such as in packing, aerospace, and protective/impact fields [1]. In general, apart from being lightweight, it is desirable that these materials/structures should exhibit both high specific strength and excellent absorption energy properties [2]. To create such structures with those desired properties, the use of only Ashby's chart for materials selection is not sufficient, since it does not take into account important topological features [3], which is also an important parameter to be considered when designing new or enhanced light-weight materials/structures [4].

In nature, mussel byssus is one of the best examples of gradient structured material. It consists of alternating soft and stiff layers. These topological concepts have been exploited to enhance the mechanical properties of metals at the microscale. This approach can be also used at the macroscale because variation in grain size, orientations and distribution hinders the stress transfer and can improve some mechanical properties [5–10].

One approach that has been used in an attempt to decrease the weight of a material is the creation of cellular non-uniform structures (such as porous foams) through a density grading [11–15]. Experimentally, these structures can be generated from bulk materials, such as polymers, metals, glass, ceramics, and composites [16]. They can be produced experimentally by, for example, a foaming process [16], and improvements in specific strength, energy absorption, and high impact tests have been observed [11–15].

Due to the intrinsic difficulties to experimentally realize such cellular griding structures, which limit their effective technological use, their fabrication using 3D printing techniques has been an effective approach to overcome some of these limitations. For example, Maskery *et al.* [17] investigated cellular lattices with graded densities printed in nylon-12 and they have shown, based on quasi-static compression experiments, that it is possible to almost double the energy absorption performance compared to corresponding uniform density structures. In another study, Bates *et al.* [15] evaluated a cellular grading honeycomb structure printed in polyurethane, and they showed that there is a considerable gain in the mechanical impact tests performance in comparison to a structure with a uniform density. These and other studies validate the effectiveness of 3D printed cellular griding structures for present and future technological applications.

It should be stressed that, in general, it is used linear graded densities in these structures, *i.e.*, the mass density varies linearly from the bottom to the top or vice-versa [13–15,17]. The obtained results for cellular griding structures showed that the mass distribution significantly influences the mechanical properties. Then, it could be interesting to evaluate/create different configurations for the graded densities and eventually determines whether it is possible to tune/calibrate the mechanical properties as a function of the mass distribution.

Recent works [18–21] have shown that mimicking the topology of structures that are found in nature is an effective way to produce new lightweight structures from nanoscale to microscale via 3D printing. Fully atomistic molecular dynamics and finite elements simulations have been used to create and test the mechanical properties of nanoscale/microscale structures. The structural information from these simulations are used to generated macroscale models that are 3D printed.

Interestingly, it was observed from the simulations and experimental tests that some of the mechanical behavior (under compressive or tensile tests, for example) are scale-independent. These results show this approach (using atomic models to create 3D print macroscale ones) is an effective way to create new macrostructures with engineered and/or enhanced mechanical properties [18–21].

In this work, based on the ideas of cellular griding structures, the gradient of grains in metals, and atomistic models from MD simulations, we proposed/created new nanostructures with a radial griding density as models for new structural architectures for 3D printing. These structures were created based on the schwarzites [18,22,23] and schwarzynes [24] classes of carbon-based materials. The unit cell of these structures are used to represent the cellular building block/grain with different pore sizes allowing to create complex architectures with different mass distributions. These new carbon-based molecular radial griding density structures can have mass densities as lower as 1.6 g/cm$^3$ and can even outperform their 'parent' schwarzites in terms of energy absorption and elastic properties. The mechanical behavior of the atomic and 3D printed models exhibit an excellent qualitative agreement, even surpassing what was observed for other 3D printed families, such as schwarzites [18] and cylinder nets [20,25,26], for example. We believe that our proposed architectures, which are low-cost, scalable, and easily fabricated, could be beneficial for several technological applications at the macroscale.

2. Methodology

2.1. Obtaining the molecular models of gradient structures

To create our structures mimicking a cellular griding, which we will call here a structure with a gradient of pores, we used as building blocks schwarzite P688 [18,22,23], and some of its corresponding schwarzynes [24] motifs. In Figure 1(a) it is presented the porous unit cells of these structures. The basic structure is the P688 Schwarzite, which we will consider as the unit of the small pore (SP) structures. According to our previous work [24], by adding acetylene groups between the benzene rings of P688, we can create a family of a new carbon-based family called schwarzynes. As can be seen in Figure 1(a), we used two of these new structures: the first one, in which it was added acetylene groups between the benzene rings of P688 along the y and z directions, and the second, where we added the acetylene groups between the benzene rings of P688 along x, y, and z directions. In comparison to their 'parent' P688 structure, we will refer to these structures as having intermediate (IP) and big pores (BP), respectively. The sizes of the SP, IP, and BP unit cells are 6.4 x 6.4 x 6.4 Å, 6.4 x 9.0 x 9.0 Å, and 9.0 x 9.0 x 9.0 Å, respectively. As can be seen from Figure 1, SP and BP are symmetrical along **x**, **y**, and **z** directions, and the IP is symmetrical along **y** and **z**, but not along the **x** one (see the used orthogonal axes shown in Figure 1).

Before creating the gradient structures, we initially built 3 (finite) structures based on SP, IP, and BP (the three first structures shown in Figure 1(b)) unit cells. These structures were created by replicating the unit cells to create an almost perfect cubic structure: **7 x 7 x 7** (size 59.2 x 59.2 x 59.2 Å, 16464 atoms); **7 x 5 x 5** (52.7 x 50.6 x 50.6 Å, 11200 atoms); and **5 x 5 x 5** (51.2 x 51.2 x 51.2 Å, 9000 atoms) for SP, IP, and

BP, respectively. These structures were built to contrast their behavior with the gradient ones, which are based on the porous unit cells of SP, IP, and BP.

Based on the SP, IP, and BP unit cells, we created two porous gradient (finite) structures, in which we used an idea of a radial gradient, *i.e.*, we started with a pore gradient from the center to the edges of the structure. Subsequently, we created two almost cubic structures in which one starts with a small pore at the center to the big pore at the edges, and another with a big pore at the center to the small pores at the edges. We will refer to these gradient structures as small-to-big (S2B) and big-to-small (B2S). The S2B and B2S structures are presented in Figure 1(b). To create the S2B structure, we started with a 4 SP square along the **xy** plane and we added 4 IP at each edge of it, forming a "+" configuration. Then, to fill the missing parts of this "+" configuration and form a square, we added 4 BP more at each corner. With this gradient structure 'motif', we repeated it 5 times along the **z**-axis (see the used orthogonal axes shown in Figure 1), resulting in the S2B structure with a size of 55.5 x 55.5 x 55.5 Å and containing 12000 carbon atoms. In Figure S1 of the Supplementary Materials is presented each step to create the S2B structure. Notice that to make a repetition of 5 times along the **z**-axis, it was necessary to add 16 acetylene groups between each 4 SP (central part) that is replicated along the **z**-axis to match with the repetition of the lateral components. The S2B structure was created in a form that each carbon atom makes 4 bonds, thus fully satisfying its valence.

As for the B2S, we started with 4 BP along the **xy** plane and we added 4 IP at each edge of it, forming a "+" configuration. To fill the missing parts of this "+" configuration and to form a square, it was added 4 SP at each corner. Then, using this structural 'motif', we repeated it 4 times along the **z**-axis (see the used orthogonal axes shown in Figure 1), resulting in the B2S structure with a size of 50.3 x 50.3 x 50.3 Å and 11040 carbon atoms. In Figure S2 of the Supplementary Materials, we present each step to create the B2S structure. Similar to what was done for the S2B structure, it was also necessary to add 16 acetylene groups between each 4 SP along the **z**-axis to match with the repetition of the lateral components. Here, we also make sure that each carbon atom makes 4 bonds. It is important to notice that S2B and B2S final structures are symmetrical along the **x** and **y** directions, but not along the **z** one.

After building these five structures (SP, IP, BP, S2B, and B2S), we performed an energy minimization procedure (molecular mechanics simulation (MM)) using a conjugated gradient technique (CG) [27,28] in order to remove any residual structural stress. We used a convergence tolerance of 0.001 kcal/mol and 0.5 kcal/mol/Å for energy and force, respectively. Then, these structures were thermal equilibrated using molecular dynamics simulation (MD) at 10K for 400ps in an NVT ensemble. All structures remained stable, and these results will be presented in the Results and Discussions section. All the energy minimizations and thermal equilibrations were carried out with the AIREBO force field [29], as implemented in the computational package LAMMPS [30]. We used a Nosé-Hoover thermostat and a time step of 0.1 fs. This methodology has been successfully used to study other porous carbon-based nanostructures [19,20,25,26].

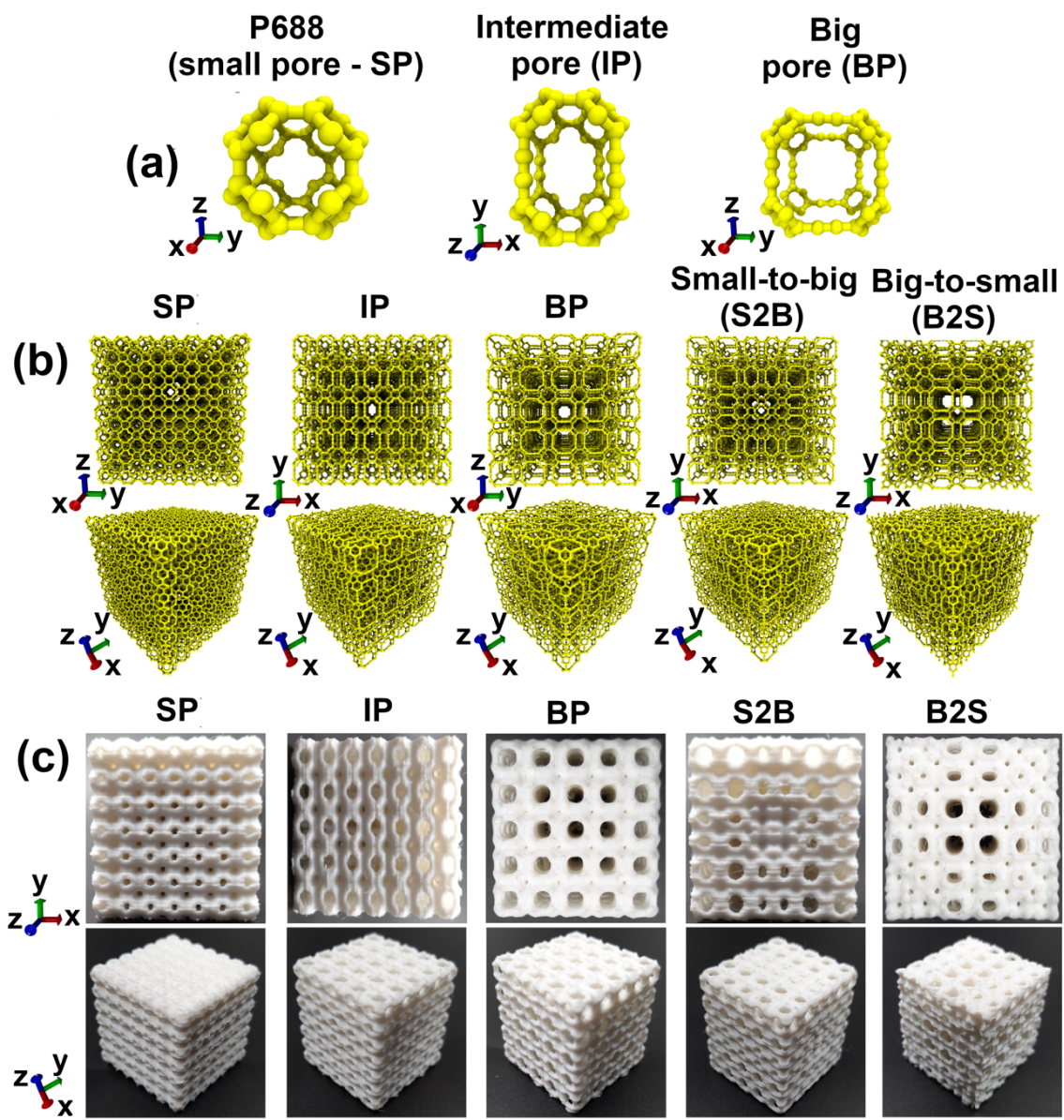

**Figure 1.** (a) Building blocks used to create the molecular gradient structures (see text for discussions). (b) Optimized structural models of the single pore (SP), intermediate pore (IP), big pore (BP), small-to-big (S2B), and big-to-small (B2S) molecular structures, respectively. (c) 3D printed models of SP, IP, BP, S2B, and B2S.

### 2.2. Preparing and printing the gradient 3D structures

After obtaining the optimized SP, IP, BP, S2B, and B2S molecular structures from MM/MD simulations, we exported them to a stereolithography file (STL) with the VMD software [31] to be 3D printed for mechanical properties evaluations.

All the structures were resized to 3 x 3 x 3 cm in the Flashprint 4.1.0 software and the converted .stl file was formatted to G-Code one. The isometric view of the structures after 3D printing is shown in Figure 1(c). The G-Code file was transferred to Filament Deposition Modeling-based Flashforge adventure 3 printer via Wi-Fi network. Solid PLA filament was fed into the extruder via guiding pipes and the filament was pulled from the spool and pushed towards the extruder with the help of a stepper motor. The extruder is heated at 210°C and the printing bed is kept at 50°C throughout the printing to avoid rapid thermal contraction. The heated nozzle has a 0.4 mm diameter and paper tape is used as a substrate for better adhesion of the first layer and for easy removal of the structures. The fill density was kept at 100% and standard resolution (single layer thickness-200 $\mu$m). The print speed and print head travel speeds were 55 mm/s and 75 mm/s, respectively. During printing, some printed parts of the structures solidify at room temperature. The solid PLA filament was supplied by Flashforge 3DTechnology Co. Ltd., Zhejiang, China. It has a 1.75 mm standard diameter with ±0.1 mm tolerance and 1.24 g/cm3 density. The solid PLA filament melts between 190°C to 220°C.

### 2.3. Compressive mechanical tests

In order to evaluate the mechanical performance of the SP, IP, BP, S2B, and B2S structures, we carried out fully atomistic simulations and experimental compressive tests.

The compressive tests for the molecular structures were performed with MD simulations. Using the energy minimized and thermal equilibrated (at 10K) structures, we placed each one of them between two fixed 12-6 Lennard-Jones walls, in which the top one is movable and the bottom one is kept fixed. In order to avoid the molecular structures slipping during the compression tests, we kept frozen the carbon atoms of the ~3 Å of the base close to the fixed wall. To make the compression, the top wall is moved towards the structure to deform it at a constant strain rate of $-10^{-6}fs^{-1}$. As the structures SP and BP are symmetrical along the **x**, **y**, and **z** directions, the compressive tests of them were performed only along the **z**-direction (see the used orthogonal axes shown in Figure 1). As IP, S2B, and B2S are symmetrical along the **x** and **y** directions, but not along the **z** one, the compressive tests were performed along the **x** and **z** directions (see the adopted orthogonal axes shown in Figure 1). The virial stress tensor component and the engineering strain at each compressive direction were used to obtain the stress-strain curves. In order to decrease the thermal contributions in the deformation tests, the compressive simulations were performed at 10 K with an NVT ensemble. To analyse where the stress is spatially accumulated during the compression tests and to obtain insights into the fracture dynamics of the studied structures, the local stress distribution was evaluated through the second invariant of the deviatoric stress tensor, *i.e.*, the von Mises stress [32]. As mentioned above, all MD simulations were carried out with the computational package LAMMPS [30], using the AIREBO reactive force field [29] and a timestep of 0.1fs.

The experimental compressive tests on the as-printed structure were carried out using UTM SHIMADZU (AG 500G) according to the adopted orthogonal **xyz** axes as a reference (see Figure 1). A constant rate of 1 mm/min was used for all the structures. To obtain further insights on the deformation mechanisms of the porous structures snapshots were captured during the tests at 30 fps. Similar to the theoretical compression tests, we performed the experimental ones along the **z**-direction for SP and BP, and for **x** and **z** ones for IP, S2B, and B2S.

## 3. Results and discussions

### 3.1. Simulation results

After the energy minimization and thermal equilibration at 10K of the SP, IP, BP, S2B, and B2S structures, the resulting mass density of each one was: 2100.0, 1536.0, 1279.0, 1421.0, and 1528.0 kg/m$^3$, respectively. Based on these mass density values, we can notice that our newly proposed 3D structures are lighter than other well-known carbon-based materials, such as graphene (2267 kg/m$^3$ [33,34]) and diamond (3510 kg/m$^3$ [33]). The size of these structures did not change significantly compared to the values of their initial configurations presented in the Methodology section. As expected, the insertion of the acetylene groups in the SP structure to form the IP and BP decreases the mass density of the final structure.

In Figure 2(a) we present the potential energy per atom during the elapsed time of thermal equilibration at 10K, which confirms the structure stability. As can be seen, the SP (-7.41 eV/atom) is the most stable one, followed by B2S (-7.10 eV/atom), IP (-7.05 eV/atom), S2B (-7.01 eV/atom), and BP (-6.93 eV/atom), respectively. If we compare these values of the potential energy per atom with the estimated cohesive energy of some of the well-known carbon-based materials, such as diamond (7.58 eV/atom), graphene (7.90 eV/atom), and γ-graphyne (6.76 eV/atom) [35], our proposed carbon-based structures are in the same energy range.

Further detailed information about the mechanical properties of SP, IP, BP, S2B, and B2S are presented in Figure S3 and Table S1 of the Supplementary Materials. As our structures have a different number of carbon atoms and their sizes are slightly different, in order to compare them, it is more meaningful to evaluate the specific mechanical properties, *i.e.*, the mechanical properties per mass density. These results are presented in Figures 2(b) and 2(c) and Table 1.

In Figures 2(b) we present the specific stress-strain curves for the compressive tests performed for SP, IP, and BP. As we can see, the insertion of the acetylene groups in the SP structure significantly affects its mechanical properties. The SP-specific yield strength and yield strains are 23.5 MPa/kg/m$^3$ and 41.4 %, respectively. As for IP, the compression along the **x**-direction (the perpendicular direction to the acetylenic groups) results in a similar specific yield strength (19.0 MPa/kg/m$^3$) and yield stain (40.7 %) values. These similar results can be attributed to the fact that topologically the IP **x**- direction is similar to SP one, they do not have acetylenic groups. For the compression along the **z**-direction (the parallel direction to the acetylene groups), the IP corresponding values decrease, presenting a specific yield strength and yield stain of 3.7 MPa/kg/m$^3$ and 15.5 %, respectively. The BP structure presents similar specific yield strength (3.8 MPa/kg/m$^3$) and yield stain (15.7 %) values. This can again be also attributed to the similarity between the IP and BP **z**-directions of IP with respect to the presence of parallel acetylenic groups along the compressive direction.

In Figure 2(c) we present the specific stress-strain curves for the compressive tests performed for S2B and B2S; we also present in this figure the SP corresponding data for comparison. The S2B presents a specific yield strength and yield strain of 4.4 MPa/kg/m$^3$ and 21.4 % for compression along the **x**-direction, and 4.7 MPa/kg/m$^3$ and 23.7% for the **z**-one. As for B2S, the corresponding values are 4.4 MPa/kg/m$^3$ and 20.8% for the compression along the **x**-direction, and 4.5 MPa/kg/m$^3$ and 21.2 % for the **z-**one. As we can see, these values are close to the ones for IP-z and BP (the shape of their corresponding specific stress-strain curves is also similar). These results suggest that although S2B and B2S contain SP motifs in their topology, apparently these properties (yield strength and yield strain) are dominated by the IP and BP building blocks. From the results discussed above, we can conclude that adding the acetylenic groups into SP and/or creating a gradient pattern does not result in any significant gain in the yield strength and yield strain, as can be observed in Figure 2(d).

Comparing the specific Young's modulus values (see Figure 2(d) and Table 1), we have 25.5, 24.5, 32.0, 36.2, 28.6, 32.1, 27.4, and 31.43 MPa/kg/m$^3$, respectively, for SP, IP-x, IP-z, BP, S2B-x, S2B-z, B2S-x, and B2S-z. The BP, S2B-z, and B2S-z have the highest specific Young's modulus, and, interestingly, they have similar mass densities, which are in between the ones of SP and BP. This result suggests that the geometrical design proposed here can result in structures enhanced specific stiffness but maintaining a relatively low mass density, which would make them good candidates for applications requiring lightweight materials with a strong and resilient structural framework.

As for the energy absorption behavior, we present in Table 1 the values for the specific modulus of resilience. This quantity is obtained from the integral of the specific stress-strain curve up to the limit of the specific yield strength. As we can see, SP and IP-x present the highest specific modulus of resilience, 235.6 and 241.8 MJ/kg/m$^3$, respectively. This is expected since the compressive tests performed for SP and IP-x show the largest yield strain and highest specific yield strength among all investigated structures. For the other structures, we have the following order (from the highest to lowest) of the specific modulus of resilience: S2B-z (65.8 MJ/kg/m$^3$), B2S-z (57.4 MJ/kg/m$^3$), S2B-x (54.8 MJ/kg/m$^3$), B2S-x (45.7 MJ/kg/m$^3$), BP (32.9 MJ/kg/m$^3$), and IP-z (29.9 MJ/kg/m$^3$). As these structures have a smaller yield strain and low specific yield strengths in comparison to SP and IP-x, these results are expected.

However, in practical applications, the structures will not be subjected to very high strain levels. Then, if we consider a regime of low deformations, i. e., up to 15% of strain, we can conclude that SP and IP-x will not be the best energy absorbers. In Figure 2(d) and Table 1, we present the values of the specific energy absorption until 15% of strain (integral of the specific stress-strain curve, from zero up to this strain level). In a low strain regime, we obtained the following order of the energy absorption (from the highest to lowest): BP (36.10 MJ/kg/m$^3$), S2B-z (32.87 MJ/kg/m$^3$), IP-z (32.81 MJ/kg/m$^3$), B2S-z (32.12 MJ/kg/m$^3$), S2B-x (30.19 MJ/kg/m$^3$), B2S-x (28.75 MJ/kg/m$^3$), SP (26.80 MJ/kg/m$^3$), and IP-x (26.12 MJ/kg/m$^3$). We can see that the BP, S2B-z, and B2S-z will be the best energy absorbers to be used in mechanical applications when it is necessary to take into account the resulting weight of the resulting structure. According to these results, we can notice that the use of a structure composed of a gradient of pores is beneficial to improve energy absorption at low strain regimes.

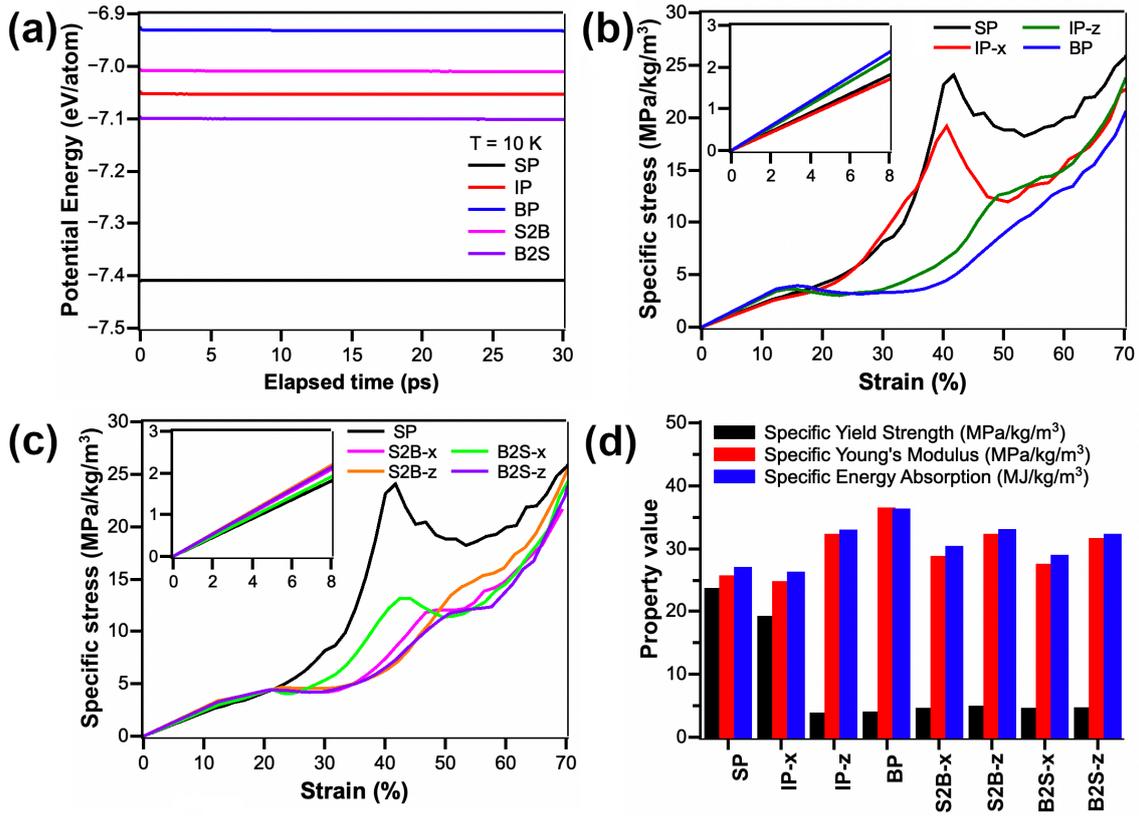

Figure 2. (a) Potential energy per atom as a function of the elapsed time of thermal equilibration at 10 K. Estimated specific stress-strain curve for (a) SP, IP-x, IP-z, and BP, and (c) SP, S2B-x, S2B-z, B2S-x, and B2S-z. The insets in (b) and (c) show a zoomed of the region of the specific stress-strain curves up to 8% of strain. (d) Estimated specific yield strength, specific Young's modulus, and specific energy absorption up to 15% of the strain of the investigated structures.

Table 1. Some specific mechanical properties of the investigated structures obtained from the compressive tests. Results from MD simulations.

| Property | SP | IP-x | IP-z | BP | S2B-x | S2B-z | B2S-x | B2S-z |
|---|---|---|---|---|---|---|---|---|
| Mass Density (kg/m$^3$) | 2100.0 | 1536.0 | 1536.0 | 1279.0 | 1421.0 | 1421.0 | 1598.0 | 1598.0 |
| Specific Yield Strength (MPa/kg/m$^3$) | 23.5 | 19.0 | 3.7 | 3.8 | 4.4 | 4.8 | 4.4 | 4.5 |
| **Specific Young's Modulus (MPa/kg/m³)** | 25.5 | 24.5 | 32.1 | 36.3 | 28.6 | 32.1 | 27.4 | 31.4 |
| **Specific Modulus of Resilience (MJ/kg/m³)** | 235.6 | 241.8 | 29.9 | 32.9 | 54.8 | 65.8 | 45.7 | 57.4 |
| **Specific Energy Absorption until 15% of Stain (MJ/kg/m³)** | 26.80 | 26.12 | 32.81 | 36.10 | 30.19 | 32.87 | 28.75 | 32.12 |
| **Yield Strain (%)** | 41.4 | 40.7 | 15.5 | 15.8 | 21.4 | 23.8 | 20.8 | 21.2 |

To better understand the deformation process and structural failure (fracture) of our proposed structures, we calculated the von Mises stress distribution per atom at selected strain levels. The results for SP, IP (along the **x** and **z** directions), and BP are presented in Figure 3. In this Figure, we also show the central unit pore of each structure along with their stress per atom distribution. It should be stressed that the values at the bottom of each structure remain minimum, independently of the strain level, because the atoms in this region are constrained during the compressive test (see section 2.3). For a better understanding of the whole process, see videos SV1 to SV4 in the Supplementary Materials.

As expected, the stress accumulation occurs mainly along the compressive directions. For the SP, as shown in Figure 3(a) (and also in video S1), the stress is almost equally distributed on all atoms, with a slightly higher accumulation in the hexagonal rings. From the analysis of the central unit pore, we can notice that increasing the strain, the hexagonal rings rotate and tend to become parallel along the x-direction and coplanar along the y-direction. This results in a decrease in the distance between the rotating rings, which will increase the electrostatic interactions, contributing to the increase in the stress level in these regions. The failure of the SP is related to the pore collapse at ~41.4 % of strain, in which the rotations of the hexagonal rings eventually result in some of the covalent bonds perpendicular to the compressive direction to break. After the yield strain, only structural densification is observed.

In Figures 3(b) and 3(c), we present the deformation process of the IP when compressed along the x and z-directions (see also the videos SV2 and SV3). For the compression along the x-direction, we can see that the most stressed regions are the hexagonal rings. Similar to the SP case, the hexagonal rings tend to become parallel to the x-direction when the strain is increased, forcing the hexagonal rings to come closer, which will increase the electrostatic interaction among them. As can be seen, the initial behavior of IP compressed along the x-direction is similar to the SP one. However, it is clear that the presence of the acetylenic groups in the direction perpendicular to the compressive one significantly alters the fracture dynamics. When the pores collapse, the deformation of acetylene groups allows the collapsed parts of the pores to slide, which causes a stress release, and the structure warps (bent/twist). In this way, the IP elastic regime along the x-direction is reduced in comparison to the SP one. As for the compression along the z-direction, we observed a different behavior, because the acetylene groups are parallel to this direction. The presence of the acetylene groups makes the pores more flexible along the z-directions, and they can stand less stress because there is no constrain to prevent them to buckle. Increasing the strain, the pores will increasingly bend until their collapse, causing the whole structure to warp early than for the compression along the x-direction.

In the BP case, as can be seen in Figure 3(d) (see also the video SV4), the presence of the acetylenic groups along the three directions allows greater structural flexibility during the compressive test. In Figure 2(b), we can observe that the elastic regimes of BP and IP-z are quite similar, and this is due to the similarity of their initial deformation process. As can be seen in Figure 3(d), the most stressed regions are the acetylene groups parallel to the compressive direction, and, similarly to the IP-z case, there is no constrain to prevent them to buckle when the strain is increased. Then, the continuous bending of the acetylene groups will cause the collapse of their pores and the whole structure to warp. Due to this deformation process similarity, BP and IP-z yield strain values are almost equal, around 15%.

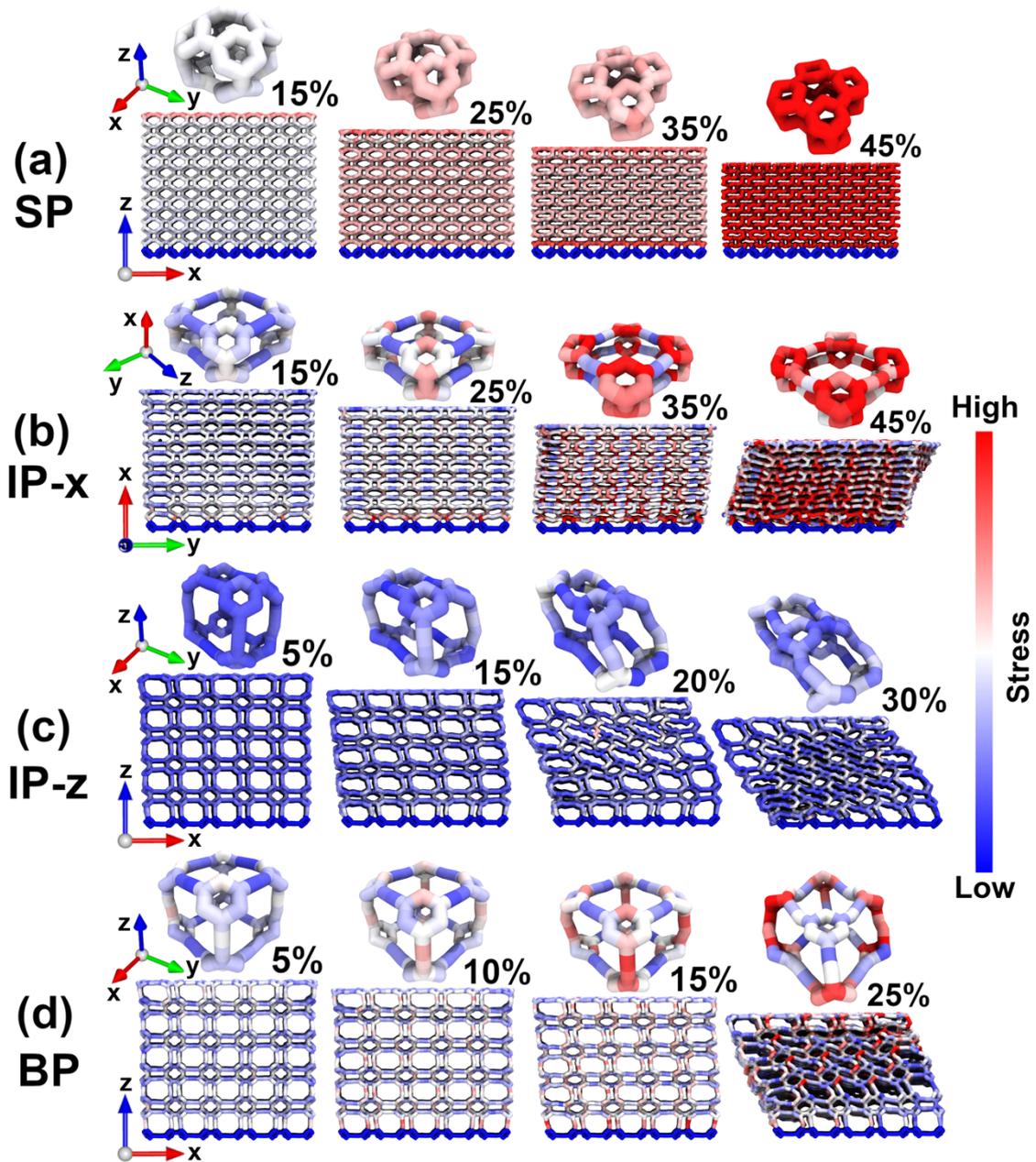

**Figure 3**. Representative MD snapshots of the stress distribution during the compressive test performed for SP, IP (**x** and **z** directions), and BP at selected strain levels.

In Figure 4 we present the (von Mises) stress distribution per atom at some selected strain levels for S2B and B2S for the compressive test performed for the x and z-directions. In the Supplementary Videos SV4 to SV8, it is presented the entire compressive tests for these structures (x and y-directions). Similar to what can be seen in Figure 3, the stress per atom distribution at the bottom of each structure in Figure 4 remains minimum independently of the strain level because the atoms in this region are constrained during the compressive tests (see section 2.3). For these structures, it is expected that the deformation process will behave as a combination of the behavior of the building blocks used to create S2B and B2S, *i.e.*, SP, IP, and BP. In Figure 4(a) it is presented representative MD snapshots of the S2B compressive test for the x-direction for selected strain levels (see also the video SV5). As can be seen, in this case when we start the compression, the stress initially accumulates in the IP and BP pore units, more specifically in the acetylene groups, *i.e.*, the surrounding regions of the SP central cluster. Increasing the strain (>20%), the stress can accumulate in this central part. This indicates, as expected, that using a structure with a gradient of pore size promotes a non-uniform stress distribution along the structure, which will influence its mechanical behavior. As seen before, the strain necessary to permanently deform the SP is higher than for IP-z and BP cases than the failure of the S2B in the x-direction is related to the collapse of the IP and/or BP pores (see the structures subjected to 30 and 35% of strain in Figure 3(a)). The failure is more pronounced in the top and bottom of the structure, which contains only the IP and BP pores. Due to this, the structure starts to warp after these pore collapse. The middle of the structure remains almost unchanged due to the intact SP central cluster, in which the stress accumulated there was not enough to deform it. Although the structural failure is due to the collapse of the IP and/or BP pores, the final yield strength is higher than that obtained for the compression of the structures composed only of these pores (see Table 1), but lower than the pure SP. This indicates that the gradient structures present an intermediary yield strain compared to their pure constituents.

In Figure 4(b) we present MD snapshots of the S2B compressive test along the z-direction for some selected strain levels (see also the video SV6). In this case, we can see that the stress starts to accumulate in the central part of the structure (up to 20% of strain), which is composed of SP pores. Then, increasing the strain (>20%), the stress is distributed to the surrounding parts, which are composed of IP and BP pores. This is also an indication of the non-uniformity of the stress accumulation along the structure. When the stress is high around the pores, IP and BP start to collapse, leading to structural failure and warp. As the surrounding will not offer structural resistance to the central part anymore, it will be pulled by the collapsed surrounding and warp together with the whole structure. As is more difficult to deform the central part than the surroundings, the yield strain for compression along the z-direction is slightly higher than for the x-direction (23.8% in comparison to 21.4%).

As for B2S, we present in Figures 4(c) and 4(d), representative MD snapshots of the structure at different strain levels from the compressive tests performed for the x and z-directions (see also videos SV7 and SV8). From these figures, it is also clear that the stress accumulation during the compression is not uniform. As can be seen in Figure 4(c), during the compression along the x-direction, the stress is higher at the SP pores located in the top and bottom of the structure (<10% of strain), and then this stress is distributed through the middle part (≥10% of strain), composed of IP and BP pores. When the stress becomes more accumulated in the middle part than the top and bottom ones (>20% of stress), the structure will fail due to the collapse of the middle pores. For the compression along the z-direction, which is shown in Figure 4(d), at the beginning, the stress will be more accumulated at the surroundings of the structure, composed of the SP and IP pores (up to 20% of strain). As the central part composed of the BP pores does not offer enough compression resistance compared to SP and IP ones, it will start to deform due to the pulling/stretching caused by the surroundings, leading the structure to fail. The final B2S yield strain for the x and z-directions is similar, being 20.8 and 21.2 %, respectively. Comparing these results with the ones obtained for S2B, the use of a structure in which the gradient is created from the small pores at the central part to the big pores at the edges promotes the structure to stand more strain than vice-versa.

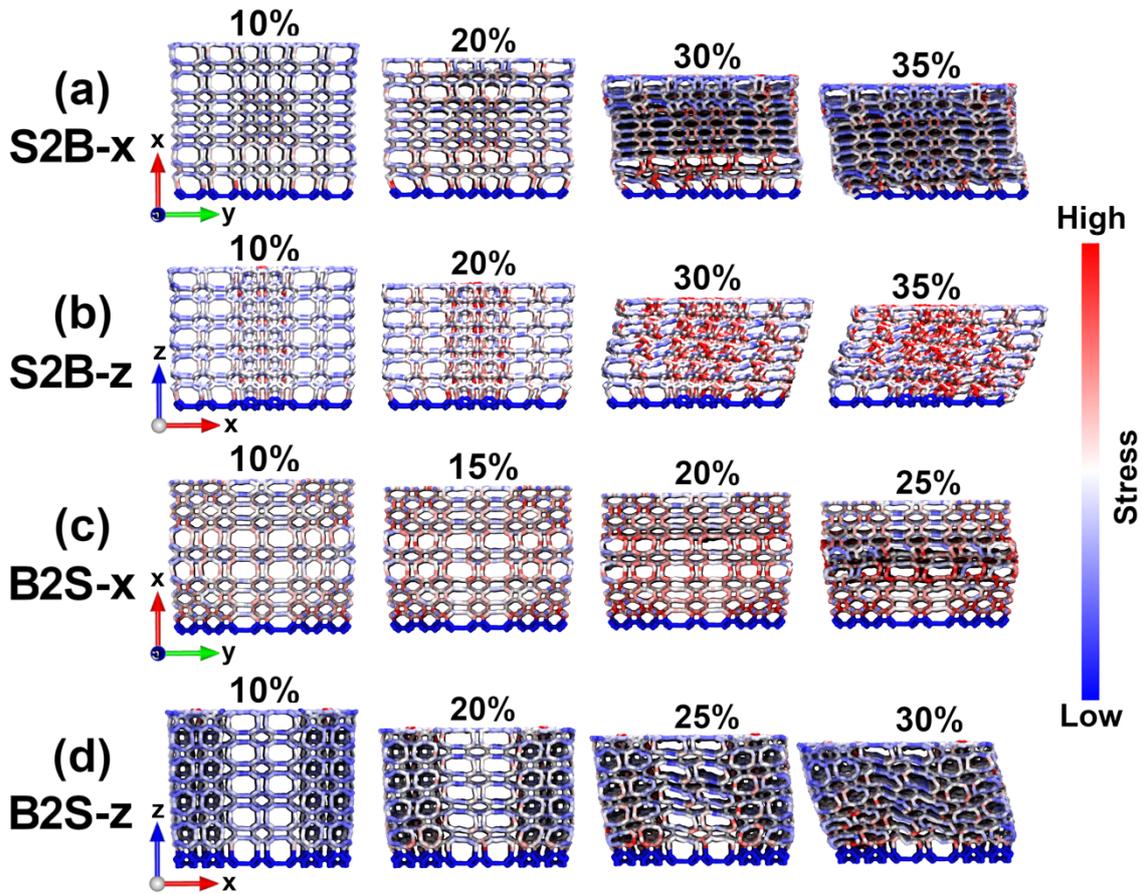

**Figure 4**. Representative MD snapshots of the stress distribution during the simulated compressive tests performed for S2B and B2S (x and z-directions) at selected strain levels.

### 3.2. Experimental results

The as-printed porous structures have porosity in the range of 45-60%. BP, IP, SP, S2B, and B2S have 64.84%, 63.05%, 45.43%, 60.48%, and 55.79% porosity, respectively. As we emphasized for the lightweight structures discussed above, density plays an important role. BP, IP, SP, S2B and B2S have bulk densities of 435.92, 458.14, 676.66, 490, and 548.14 kg/m$^3$, respectively. BP and SP have similar pore geometry along the x, y, and z-directions therefore we have tested these structures only along the z-direction. As the IP x and y-directions are similar to SP ones, we will also test them only along the z-direction. Hierarchical structures like S2B and B2S have different pore geometry along the x and z-directions therefore to study the directionality/topological effect on the mechanical behaviour we have tested them for both the x and z-directions.

In **Figures 5(a) and (b)** we present the experimentally obtained specific compressive stress-stress curves of the porous structures. IP has a higher slope than SP and BP. Initially, SP has a lower slope than BP but at 2.2% strain, BP has a lower strain than SP. In the case of the hierarchical structures, firstly SP has a higher slope but after 2.8% and 3.8% strain, B2S-x and B2S-z, respectively, overpasses SP and exhibit excellent mechanical performance. Amongst all the structures, S2B-z presents the lowest slope. In **Figure S4** we present the experimental compressive stress-stress curve for all the porous structures. Compressive strength is inversely proportional to pore size. For the initial deformations (up to 5%,), the stress-strain curves show that SP has a higher slope than IP and BP (**Figure S4(a)**). Amongst all the hierarchical structures B2S-x presents the highest slope and S2B-z the lowest one (**Figure S4(b)**).

To directly compare all the lightweight porous structures, we normalized their densities and calculated the corresponding specific Young's modulus, specific yield strength, and specific energy absorption (toughness). In **Table 2**, we present the data for some mechanical properties obtained from **Figures 5(a)** and **5(b)**. From **Figure 5(c)** we can see that the specific yield strength values increase with the decrease of pore size, from BP to IP, but afterward, they decrease from IP to SP. The hierarchical structure S2B-z exhibits the lowest specific yield strength but B2S-z has higher specific yield strength than BP, IP, and SP. This behavior can be explained due to the arrangement of the hierarchical pores which can more efficiently constrain the deformations than the periodic arrangement of the porous structures. For the x-direction, both B2S and S2B structures show higher specific yield strength than for the z-direction, for the same reasons discussed above. From **Figure 5(c)** we can see that Young's modulus values are inversely proportional to pore size, SP has a higher value than IP and BP. The yield strength values also exhibit similar behavior for the hierarchical structures. B2S-x and B2S-z have higher Young's modulus values than S2B-x and S2B-z. Amongst all the structures B2S-x has the highest specific Young's modulus value.

Structural energy absorption capability is of critical importance in some applications, such as in the automotive industries specifically for the car crumple areas. From **Figure 5(c)**, we can see that for our investigated structures the specific energy absorption increases with the decrease in pore size and then decreases, *e.g.*, IP has higher specific energy absorption than BP and SP structures. Hierarchical structures like B2S-z have higher specific energy absorption than BP, IP, and SP. B2S-x and S2B-x both have the highest specific energy absorption values amongst all the porous structures.

It should be stressed that in the FDM-based printing technique, the structural models are printed layer-by-layer manner therefore the resulting printed structures have intrinsically anisotropic properties. While studying the deformation behavior of the porous structures this anisotropic can affect the results. To overcome this issue we have printed structures in such a way that the load is always transverse to the printed layers so that anisotropic will not affect the final test results.

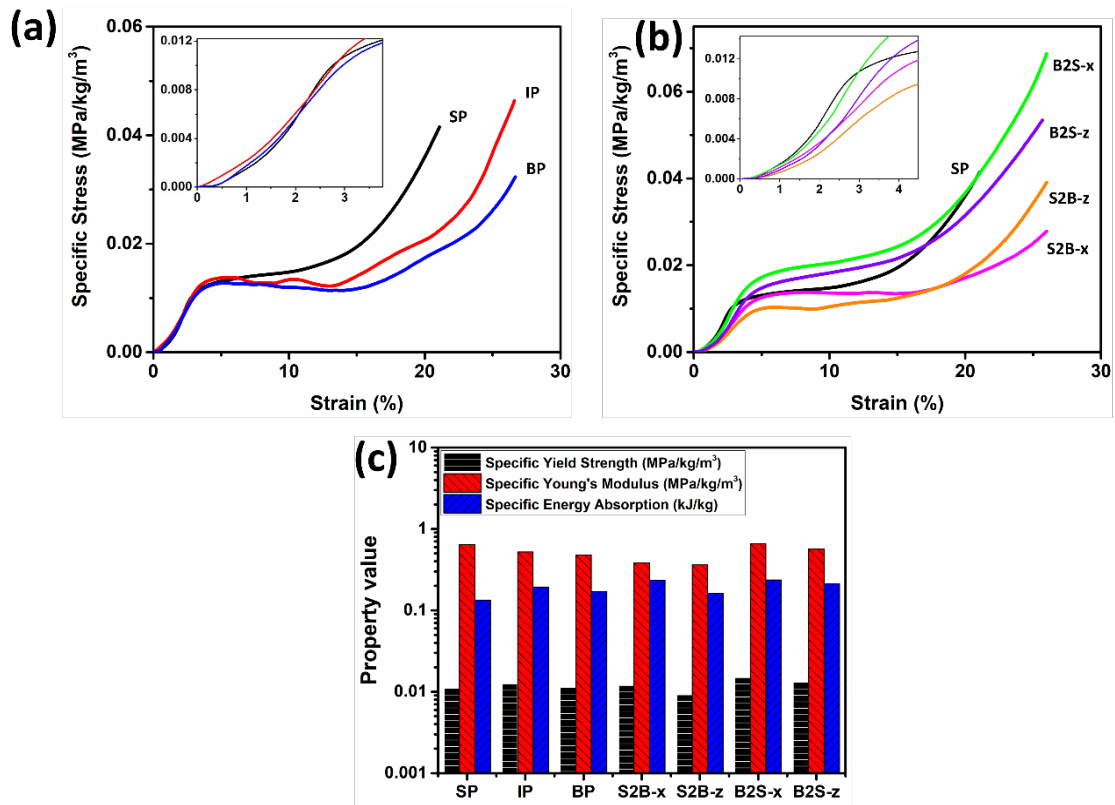

**Figure 5.** Experimentally obtained specific stress-strain curves for (a) SP, IP, and BP, and; (c) SP, S2B-x, S2B-z, B2S-x, and B2S-z. The insets in (a) and (b) show the initial specific stress-strain curve up to 4% of strain. (d) Experimentally obtained values for specific yield strength, specific Young's modulus, and specific energy absorption, respectively, of the investigated structures.

**Table 2.** Experimental specific mechanical properties of the studied structures obtained from the compressive testing.

| Property | SP | IP | BP | S2B-x | S2B-z | B2S-x | B2S-z |
| --- | --- | --- | --- | --- | --- | --- | --- |
| Bulk Density (kg/m³) | 676.66 | 458.14 | 435.92 | 490.00 | 490.00 | 548.14 | 548.14 |
| Specific Yield Strength (MPa/kg/m³) | 0.0110 | 0.01257 | 0.0113 | 0.0120 | 0.0092 | 0.0150 | 0.0131 |
| Specific Young's Modulus (MPa/kg/m³) | 0.6446 | 0.5196 | 0.4803 | 0.3830 | 0.3608 | 0.6630 | 0.5657 |
| Specific Energy Absorption (kJ/kg) | 0.1338 | 0.1935 | 0.1712 | 0.2333 | 0.1625 | 0.2358 | 0.2118 |

In **Figures 6** and 7, we present representative snapshots of the deformation of the porous (hierarchical and non-hierarchical) structures at selected compression strain values. The general deformation trends are in good agreement with the ones obtained from the MD simulations. In general, we can see that as the load increases the pore shape changes (from circle to oval ones), and then the pore close, which leads to structure failure (structural collapse). We have observed similar a phenomenon for the non-hierarchical structures such as BP, IP, and SP. They experience load transfer from the top portion to the bottom one, which is deformed more rapidly than the top part.

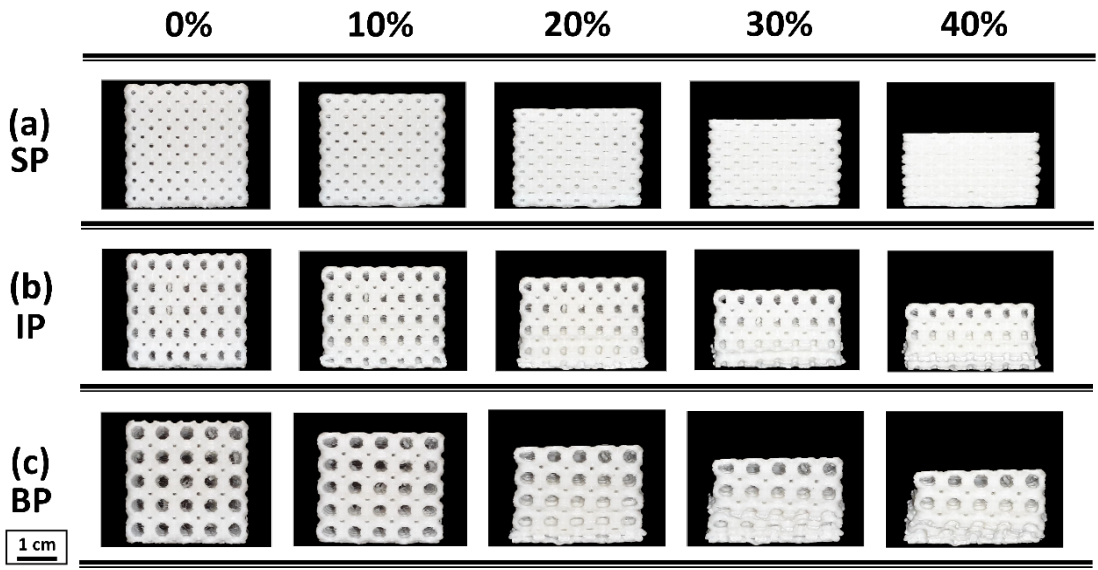

**Figure 6.** Representative snapshots at selected strain values of the stress distribution during the experimental compressive test performed on SP, IP, and BP.

As can be seen from **Figure 7,** the hierarchical structures exhibit a different deformation behavior than the non-hierarchical ones (**Figure 6)** and which can also be seen in **Figure 4** for the MD results. For the x-direction, B2S-x deforms top and bottom parts first and then the middle part, whereas for S2B-x the middle deforms first and then top and bottom ones. For the z-direction, B2S-z deforms uniformly throughout the structure, whereas for the S2B-z the bottom part deforms first. Major structural deformations were observed after 20% compression followed by densification (after 30%).

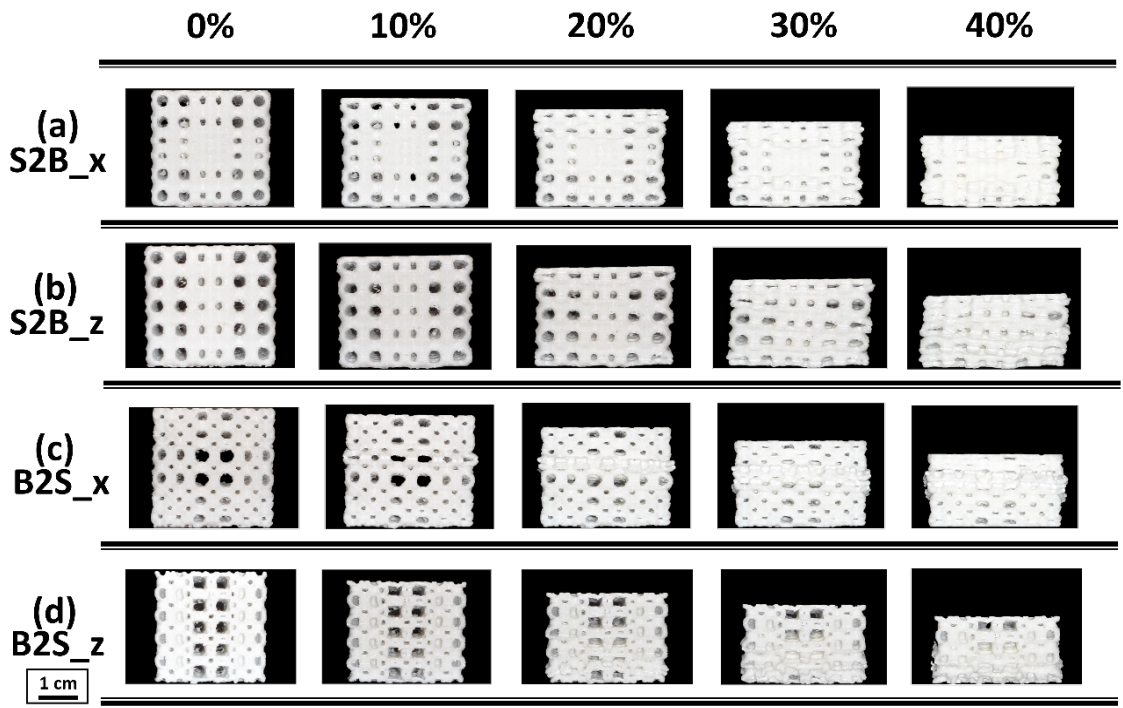

**Figure 7.** Representative snapshots at selected strain values of the stress distribution during the experimental compressive test performed on S2B and B2S (x and z-directions).

4. **Summary and Conclusions**

In this study, we have proposed a new way to create cellular griding structures, in which the mass density varies from the center to the borders, *i.e,* a radial gradient. To produce a cellular griding, our structures were based on two kinds of carbon-based structures with different pore sizes, schwarzites, and schwarzynes. These fully atomistic models were then translated into macroscale ones that were then 3D printed.

Our theoretical results from molecular dynamics simulations show that lighter and very stable structures can be obtained in comparison to other well-known carbon-based materials, such as graphene and diamond. From the mechanical point of view, the specific yield strength and yield strain of the gradient structures do not have present significant gains in comparison to structures with a uniform density. However, the geometrical design proposed here results in structures with improved specific stiffness, while presenting a relatively low mass density, which are important properties for applications that require strong and/or resilient structural frameworks in association with the minimum weight possible. Regarding the radial griding density pattern, structures in which the gradient is created from the small pores at the center to the big ones at the corners/edges allow them to stand more strain than the vice-versa.

The results from the mechanical tests carried out for the 3D printed structures show a remarkable qualitative agreement with the ones from MD simulations, which are an indication that the general trends of the mechanical behavior are scale-independent. The as-printed structures have porosity in the range of 45-60%, and the specific energy absorption values increase with the decrease of the pore sizes. In general, the gradient nature of structures affects their porosity and, consequently, their specific Young's modulus and yield strength values, which are superior in comparison to the ones from uniform density structures.

## 5. Acknowledgements

EFO and DSG thank the Brazilian agencies CNPq and FAPESP (Grants 2013/08293-7, 2016/18499-0, and 2019/07157-9) for financial support. Computational support from the Center for Computational Engineering and Sciences at Unicamp through the FAPESP/CEPID and the Center for Scientific Computing (NCC/GridUNESP) of São Paulo State University (UNESP) are also acknowledged. This study was financed in part by the Coordenação de Aperfeiçoamento de Pessoal de Nível Superior - Brasil (CAPES) –Finance Code 001.